# Kirchhoff's Forgotten Contributions to Electromagnetism: Continuity Equation versus Displacement Current

**Xavier Oriols,** *Member IEEE*

*Departament d'Enginyeria Electrònica, Universitat Autònoma de Barcelona, 08193 Barcelona, Spain*

**Robert Eisenberg,** *Life Member IEEE*

*Department of Physiology and Biophysics, Rush University Medical Center, Chicago, IL; Department of Applied Mathematics, Illinois Institute of Technology, Chicago, IL; Department of Biomedical Engineering, University of Illinois Chicago, Chicago, IL USA*

**David K. Ferry,** *Life Fellow IEEE*

*School of Electrical, Computer, and Energy Engineering, Arizona State University, Tempe, AZ 85287USA*

## I. INTRODUCTION

Coulomb, Gauss, Ampère, Weber, Faraday, Kirchhoff, Maxwell—these are giants whose work shaped our understanding of electrical circuit theory and electromagnetism. Most of them are widely recognized for their contributions to the understanding of electromagnetic phenomena in various scenarios. Kirchhoff, on the other hand, is best known for his current and voltage laws in electrical circuits and for his later work on black-body radiation from heated objects. It is generally considered that he made no major contribution to the development of electromagnetic theory.

In this sense, it is often overlooked that, in 1857, Kirchhoff published two seminal papers on the motion of electricity in wires [1,2], building upon Weber's electrodynamic theory [3]. In that work, he was the first to derive what we now call the telegrapher's equations, which describes the propagation of electromagnetic signals along a cable [4,5].

But how was Kirchhoff able to describe electromagnetic propagation as early as 1857, when the notion of displacement current—which is believed to be the essential ingredient for the propagation of electric and magnetic fields—was not introduced by Maxwell until 1861 [6,9] and fully explained later in 1865 [10]? Did Kirchhoff somehow anticipate the concept of displacement current before Maxwell? Or was his work simply incorrect because, at that time, he did not have the tools to describe the propagation of electromagnetic waves?

### A. A Wrong Consensus

If we take Wikipedia as of November 2025 [11] as a rough indicator of the general consensus on who should be credited with developing the telegrapher's equations, we see that Kirchhoff's contribution is essentially ignored. The authorship of the telegrapher's equations is commonly attributed to Heaviside [12], who in 1876 —this time after the discovery of the displacement current by Maxwell—rederived the same equations that Kirchhoff had already found in 1857 [1].

The reasoning behind the widespread dismissal of Kirchhoff's contribution appears to be [13-16]:

1. Kirchhoff based his work on Weber's electrodynamics [3], which, in turn, was based on the *primitive* notions of velocities, accelerations, and forces, instead of Maxwell's *modern* concept of fields [6-10].
2. Thus, Kirchhoff was unable to discuss the displacement current because he missed the notion of fields.
3. To conclude, Kirchhoff's result cannot be correct, since displacement current is regarded as the essential ingredient for electromagnetic propagation.

The previous argument suggests that Kirchhoff's original mistake in Refs. [1,2] was basically adhering to Weber's *outdated* ideas [3] for explaining electromagnetism without invoking the *modern* concepts of fields (i.e., the concept of displacement currents). But, there is nothing wrong—or at least nothing that invalidates the demonstration of propagation of electromagnetic signals—within Weber's framework [17].

All physical theories are valid only within a range of validity. The Weber theory is valid for classical electrons with velocities lower than the velocity of light, which is the typical scenario in metals [17]. In this paper, we show in detail that the previous consensus concerning Kirchhoff is wrong. He must be recognized as the first to discuss the propagation of electromagnetic signals in wires, even though he did not introduce the displacement current in his approach (because he did not need it).

Discrediting Kirchhoff's contribution simply because he did not use fields or the concept of displacement current is, in our view, a serious historical mistake [17-20]. This paper aims to help correct that misconception about Kirchhoff's forgotten contribution to electromagnetism. Fig. 1 shows portraits of the three main relevant figures mentioned in this paper.

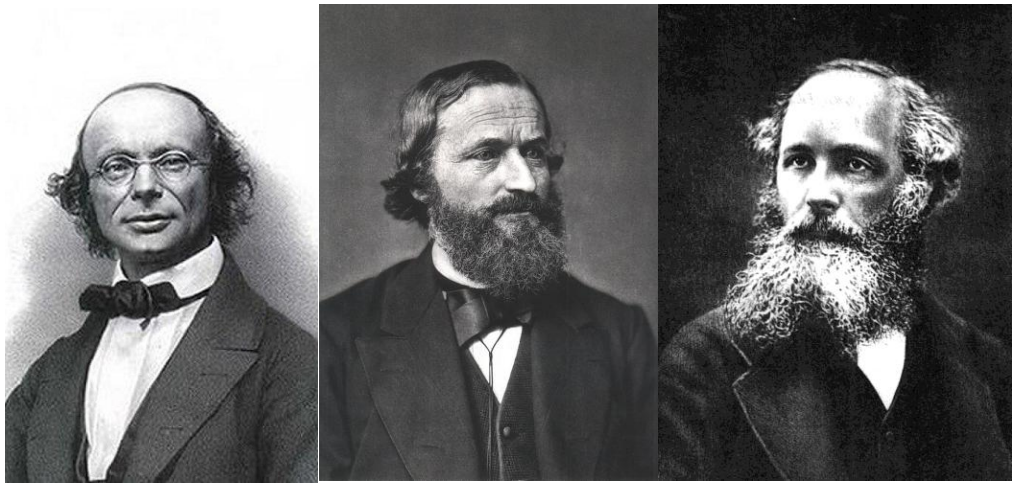

Fig. 1. *Pioneers of the electromagnetism mentioned in this paper. From left: Wilhelm Weber, Gustav Robert Kirchhoff and James Clerk Maxwell.*

## B. A New Perspective

In this paper, we argue that a new perspective on the role of the continuity equation for matter in the development of electromagnetism is needed. The continuity equation is a more fundamental concept than that of the displacement current. The dynamics of charged particles can be formulated without invoking fields, but the dynamics of fields cannot be formulated without a continuity equation for matter [15,20-24]. In modern quantum theories of electrodynamics, where electromagnetic fields are quantized, the concept of displacement current is usually ignored (it is hidden behind tons of mathematical non-commuting operators [25,26]), while the continuity equation remains an essential element in the formulation of quantum electrodynamics (see Appendix in Ref. [27]). In addition, the continuity equation is Lorentz-invariant, ensuring its validity even in relativistic scenarios [28].

By centering the discussion on the continuity equation —rather than in the displacement current— we argue that, when Maxwell introduced the displacement current in his formalism of electromagnetic fields, he introduced the continuity equation of matter as a fundamental element in describing electromagnetic phenomena. What he effectively did—without explicitly realizing it [29-32]—was to impose a restriction on the types of currents and charges that generate electromagnetic phenomena in nature; he introduced the continuity equation among the set of Maxwell's equations.

Kirchhoff was the first, in his 1857 paper [1], to introduce the continuity equation when discussing electromagnetic propagation. Although Kirchhoff employed Weber's particle-based ideas [3], he did not describe the interaction between individual particles, but rather dealt with ensembles of them expressed in terms of current and charge. Therefore, he needed a formulation of the continuity equation to conserve the number of particles locally. If one understands Maxwell's introduction of the displacement current as, in essence, the imposition of a continuity equation – what Maxwell originally called "molecular vortex" [31]– on the type of matter that can generate electromagnetic phenomena, then Kirchhoff's work predates and anticipates this important conceptual contribution. With his continuity equation, Kirchhoff was able to describe the propagation of electromagnetic signals in wires, without the need to invoke the displacement current.

## II. THE TELEGRAPHER'S EQUATIONS

The central element that we analyze in this paper to document Kirchhoff's contribution to electromagnetism is his paper written in 1857 and entitled "*On the motion of electricity in wires*" [1]. In that paper, he develops the so-called telegrapher's equations [33], showing that in a circuit of negligible resistivity, oscillating currents propagate along the wire with a velocity equal to the speed of light.

### A. Modern formulation

Before discussing Kirchhoff's work, we summarize the modern formulation of the telegrapher's equations in terms of (lumped) circuit elements as can be found in many modern textbooks [34,35]. A small section of a wire, from positions $a$ to $b = a + \Delta x$, is modelled by a capacitor $C$ (per unit length), inductor $L$ (per unit length) and a resistor $R$ (per unit length) as shown in Fig. 2 [36].

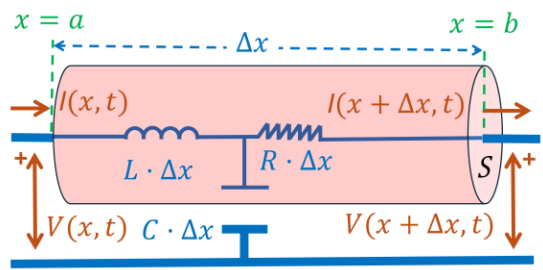

Fig. 2. *Transmission line in terms of lumped elements (in blue), as analyzed in textbooks to describe signal propagation. The cable (in orange), as analyzed by Kirchhoff in Ref. [1].*

Taking into account the current flowing into the plates of the capacitor, the current entering this section, $I(a,t)$, and leaving it, $I(b,t)$, can be written as

$$I(a,t) = I(b,t) + C\,\Delta x\, \frac{\partial V(x,t)}{\partial t}. \qquad (1)$$

It is assumed that $x \in [a,b]$ with $\Delta x \to 0$. Then, under the approximation $\frac{I(b,t)-I(a,t)}{\Delta x} \approx \frac{\partial I(x,t)}{\partial x}$, Eq. (1) becomes

$$-\frac{\partial I(x,t)}{\partial x} = C\,\frac{\partial V(x,t)}{\partial t}. \qquad (2)$$

The potential drop in the section $\Delta x$ in Fig. 2 can be written as the sum of the voltage drop in the resistor plus in the inductor

$$V(a,t) - V(b,t) = \Delta x \left( R\, I(x,t) + L\,\frac{\partial I(x,t)}{\partial t} \right). \qquad (3)$$

By assuming $\Delta x \to 0$ in Eq. (3), i.e. $\frac{V(b,t)-V(a,t)}{\Delta x} \approx \frac{\partial V(x,t)}{\partial x}$ one obtains

$$-\frac{\partial V(x,t)}{\partial x} = R\, I(x,t) + L\,\frac{\partial I(x,t)}{\partial t}. \qquad (4)$$

Taking the time derivative of Eq. (2) and the spatial derivative of Eq. (4), and equating the mixed term $\frac{\partial^2 V(x,t)}{\partial x \partial t}$, one gets

$$\frac{\partial^2 I(x,t)}{\partial x^2} = RC\,\frac{\partial I(x,t)}{\partial t} + LC\,\frac{\partial^2 I(x,t)}{\partial t^2}. \qquad (5)$$

For a wire with a low resistance, i.e., $R = 0$, Eq. (5) describes the propagation of $I(x,t)$ at a velocity $1/\sqrt{LC}$ determined by the transmission medium [34,35]. Notice that the electrons themselves do not move fast; it is the time-dependent currents that vary rapidly. Current is the movement of charge, but individual particles do not need to move at the same speed. The so-called Newton's cradle illustrates this difference: suspended balls transfer energy and momentum through elastic interactions, while the balls remain nearly stationary.

**B. Physical interpretation**

Using the relation between the voltage and the charge $Q(x,t)$ in a capacitor

$$Q(x,t) = C\Delta x\, V(x,t), \qquad (6)$$

and defining the charge density as $\rho(x,t) = Q(x,t)/(S\,\Delta x)$ and the current density $J(x,t) = I(x,t)/S$ with $S$ the cross sectional area of the wire in Fig. 2, Eq. (2) can be rewritten as

$$\frac{\partial J(x,t)}{\partial x} = -\frac{C}{S}\,\frac{\partial V(x,t)}{\partial t} = -\frac{\partial \rho(x,t)}{\partial t}, \qquad (7)$$

which can be interpreted as a charge conservation law in the region $\Delta x = \partial x$ and time $\partial t$: The electrons leaving the section $\Delta x$ in Fig. 2, $I(b,t)$, equal those entering, $I(a,t)$, minus those accumulated within that region during $\partial t$.

The term $R\, I(x,t)$ in Eq. (3) captures the idea that any instantaneous net charge different from zero at a given position in the cable, as discussed in Eq. (7), creates a Coulomb force [37] between electrons, causing them to move. Any acceleration of an electron due to Coulomb forces increases its kinetic energy, which is rapidly dissipated through energy exchange with phonons, i.e., via Joule heating [38], keeping a constant velocity [16].

The term $L\,\frac{\partial I(x,t)}{\partial t}$ in (3) represents the combined effect of Ampère's law [39] (a current flowing through the cable generates a magnetic field around it) and Faraday's law [40] (a time-varying magnetic flux induces a voltage—the electromotive force—along the cable).

By isolating the current, Eq. (4) can be rewritten as

$$I(x,t) = \frac{1}{R}\,\frac{\partial V(x,t)}{\partial x} - \frac{L}{R}\,\frac{\partial I(x,t)}{\partial t}, \qquad (8)$$

which corresponds to the generalized Ohm's law.

$$J = \sigma E \quad \text{with} \quad E = -\frac{\partial V(x,t)}{\partial x} + \frac{\partial A(x,t)}{\partial t}, \quad (9)$$

with the vector potential $A(x,t)$ proportional to the current $I(x,t)$, as happens in an inductor.

In Eq. (3), we have identified two physical mechanisms that decelerate electrons: the interaction with phonons, and the inductive effects, modeled by Ampère's and Faraday's laws. The relative importance of these two acceleration/deceleration mechanisms depends on the condition

$$L\frac{\partial I(x,t)}{\partial t} \gg RI(x,t). \quad (10)$$

The faster the time derivative of the current, the more important the inductive effects.

## C. Kirchhoff's formulation in 1857

Kirchhoff describes the transport of current in the section $\Delta x$ of the cable depicted in Fig. 2 [41]. In his paper [1], there are only four numbered equations, which are the basic results, whose combination leads him to his final telegrapher's equations. We summarize below each of these four equations.

**First equation:** The equation labelled as (1) in Ref. [1] is the "*electrostatic law of Coulomb*", where Kirchhoff establishes a relation between the charge and the potential.

$$V = 2e \ln\left(\frac{2\varepsilon}{\alpha}\right) + \int \frac{e'}{r} ds'$$
$$= 2e \ln\left(\frac{l}{\alpha}\right). \quad (11)$$

He uses the variable $s$ as our variable $x$ and $r$ as the distance between the point $x$ and any other point $x'$ in the cable. The integration is extended over the whole section of the cable of length $l$. Here, $\alpha$ is the radius of the wire and $\varepsilon$ is a small (intermediate) length used in the calculations. Finally, $e$ is the "quantity of electricity" which is the historical term to refer what today we define as charge density $\rho(x,t)$.

Expression (11) can easily be recovered from the Coulomb law assuming $l \gg \alpha$ and that only the charge near the position $x$ is relevant for the potential there [36]. Of course, Eq. (11) plays the role of a capacitance shown in Fig. 2 and mentioned in Eq. (6).

**Second equation:** The equation labelled as (2) in Ref. [1] is the most relevant equation in Kirchhoff's paper

$$i = -2\pi k\alpha^2 \left(\frac{\partial V}{\partial s} + \frac{4}{c'^2}\frac{\partial w}{\partial t}\right). \quad (12)$$

Here, $i$ is the current density (that we have named $J = I/S$), $k$ is the conductivity constant and $\frac{\partial V}{\partial s} = -E$ is the electric field. We write $c'$ as Weber's constant, linked to what today is defined as the speed of light, $c = 3 \cdot 10^8$ m/s, through $c = c'/\sqrt{2}$.

The first term, $i = -2\pi k\alpha^2 \frac{\partial V}{\partial s}$, in Eq. (12) is clearly Ohm's law, which plays the role of the resistance R in the circuit of Fig. 2. In the bottom of page 397 Kirchhoff writes [1]: "*In the case of a stationary electric current, the density of the current is equal to the product of the electromotive force, referred to the unit of quantity of electricity, and the conductivity; I will assume that the same also holds good when the current is not stationary. This assumption will be fulfilled when the forces acting upon the electricity, and which constitute the resistance, are so powerful that the time during which a particle of electricity remains in motion after the cessation of the accelerating forces, and in virtue of its inertia, may be regarded as infinitely small, even in comparison with the small space of time which comes into consideration in the case of a non-stationary electric*" Kirchhoff is acknowledging here that the transport in the wires is diffusive, not ballistic.

The second term in Eq. (12), involving $w$, requires Kirchhoff's third equation, where he defines it.

**Third equation:** The equation labelled as (3) in Ref. [1] is written as

$$w = 2i \ln\left(\frac{2\varepsilon}{\alpha}\right) + \int \frac{i'}{r} ds' \cos(\theta) \cos(\theta')$$
$$= 2i \ln\left(\frac{l}{\alpha}\right). \qquad (13)$$

Thus, $w$ in the modern formulation is just the axial component of the vector potential $A(x,t)$ which is proportional to $i(x,t)$ in a cable. Then, the second term $\frac{4}{c^2}\frac{\partial w}{\partial t}$ in Eq. (12) is just the time derivative of the vector potential, defined in Eq. (9), which gives an induced electric field due to time-dependent variation of the current [42].

The contribution of the second term of Eq. (12) can be easily recovered from Ampere's law, assuming $l \gg \alpha$ so that only the current near the position $x$ is relevant [42]. Of course, the role of this second term is just the inductor placed in Fig. 2.

Kirchhoff wrote in page 396 in the second paragraph [1]: *"We have now to form the expression for the electromotive force induced in the point under consideration, by the alteration of the intensity of the current in all portions of the wire"*. Kirchhoff acknowledged here that there are forces at a point $x$ created by the accelerations of electrons (what he named *"alteration of the intensity of the current"*) at another point $x'$. He is applying here Weber's force between charges by knowing that current is proportional to the velocity of charge and the time-derivative of the current is proportional to the acceleration of charges. He wrote [1] in page 396, *"When in the element of a conductor…the intensity of the current denoted by I' changes, an electromotive force will be induced by this change in a second element of the conductor."*

Now, it is evident that the second equation in Kirchhoff's paper, numbered here as Eq. (12), is just the modern Eqs. (8) and (9). Let us give some details on how Kirchhoff was inspired by Weber's previous work in developing Eq. (12).

Weber's Electrodynamics theory [3,17] aimed to generalize Coulomb's [37] and Ampere's [39] laws to account for the motion of charges, proposing a force law that depends not only on distance but also on relative velocity and acceleration. Two particles $q_1$ and $q_2$ at positions $\mathbf{r_1}$ and $\mathbf{r_2}$ suffer a force given by

$$\mathbf{F} = \frac{q_1 q_2}{4\pi\epsilon\epsilon_0 r^2}\left(1 - \frac{\dot{r}^2}{2c'^2} - \frac{r\ddot{r}}{2c'^2}\right)\mathbf{u_r}, \qquad (14)$$

with $r = |\mathbf{r_1} - \mathbf{r_2}|$ and $\mathbf{u_r} = (\mathbf{r_1} - \mathbf{r_2})/r$. For the cable considered here, the current $I(x,t)$ is proportional to the electrons' drift velocity. Thus, its time derivative $\partial I(x,t)/\partial t$ can be associated with the time derivative of the velocity, i.e., the acceleration of electrons given by $\ddot{r}$ in Eq. (14). Thus, Eq. (14) establishes a force between electrons due to the time derivative of the current. This new induced electromotive force contributes to the net current of the electrons in the wire as shown in Eq. (12), or in its modern version in Eq. (8). This induced electromotive force opposes the original electrons' acceleration [43]. See Ref. [42] for a more detailed explanation.

Up to here, Kirchhoff has three equations, but four unknowns: *"To the equations (1), (2), (3), between four quantities i, e, V, w, a fourth may be added."* [1]

**Fourth equation:** His fourth equation is the continuity equation written in Ref. [1] as

$$2\frac{\partial i}{\partial s} = -\frac{\partial e}{\partial t}. \qquad (15)$$

From Fig. 2, Eq. (15) means that the output current $I(b)$ at $x = b$ can be related to the input current $I(a)$ through a Taylor expansion, i.e., $I(b) = I(a) + \frac{\partial i}{\partial s}S\Delta x$ with $\partial s = \partial x$ and $I = iS$ to take into account the cross sectional area. Thus, this difference between input and output currents, multiplied by a small time interval $\partial t$, implies an increment of charge during this time given by $\frac{\partial i}{\partial s} S \Delta x \partial t = \partial e S \Delta x$ where $\partial e$ is defined as the increment of charge density in the volume $S \Delta x$ of the cable of Fig. 2. This last relation can be written as $\frac{\partial i}{\partial s} = -\frac{\partial e}{\partial t}$, meaning that nonuniformity of currents is linked to the temporal variation of charges. The factor of 2 in Eq. (15) arises from considering the positive and negative quantities of electricity at that time [44].

It seems quite reasonable to suppose that Kirchhoff's reasoning on the continuity equation in Eq. (15) was based on Weber's picture of interacting particles. The number of particles entering a region must equal the number of particles leaving it, plus those accumulating in that region, regardless of how small the region or the time interval involved.

**Final telegrapher's equations**: Once the previous four equations were established, Kirchhoff provided a rather long and tedious derivation to solve this system of 4 equations. In fact, to obtain our modern Eq. (5), one simply needs to introduce the relations between $V$ and $e$ given by Eqs. (11) into (12), introduce the relations between $w$ and $i$ given by Eqs. (13) into Eq. (12), and then take its time derivative. Finally, applying the continuity equation in Eq. (15) to express the result solely in terms of the current $i$, one obtains

$$\frac{\partial i}{\partial t} = -4\pi k \alpha^2 \ln\left(\frac{l}{\alpha}\right)\left(-2\frac{\partial^2 i}{\partial s^2} + \frac{4}{c^2}\frac{\partial^2 i}{\partial t^2}\right), \quad (16)$$

which reproduces Eq. (5) with $I = i S$.

In summary, this last equation (16) is equal to Eq. (5)—which appears in almost all electrical engineering textbooks dealing with electromagnetic transmission in cables [33,34]. Both equations are based on the same physics. In fact, taking the time and spatial derivatives of Eq. (1) in Heaviside's original work in Ref. [12] and substituting them into the time derivative of Eq. (2) in Ref. [12] yields Kirchhoff's original equation (16) once again. How can one claim that Kirchhoff's equation is wrong while Heaviside's equation is correct if they are in fact the same? The argument that Kirchhoff did not include the displacement current is meaningless, since he described particle electrodynamics without invoking electromagnetic fields, relying solely on Weber forces.

### III. IS WEBER'S THEORY RIGHT?

Historically, the first descriptions of electrodynamics were formulated as mechanical forces between particles, acting instantaneously at a distance. For example, Coulomb [37] described electrostatic interactions through direct forces between charges, while Ampère [39] extended this view to moving charges, revealing forces between electric currents. Weber [3] generalized Coulomb's law to include Ampère's formalism, thus providing the first unified instantaneous action-at-a-distance formalism for electricity and magnetism [3,17]. Faraday [40] then transformed this picture by introducing the idea of continuous "*lines of force*", suggesting that space itself mediates electrical and magnetic actions. Maxwell [6-10] finally formalized this vision mathematically, defining the electric and magnetic fields as local physical quantities whose variations propagate as electromagnetic waves. Through his four papers [6-9] leading up to the *Treatise* [45], Maxwell showed that interactions among particles can be also understood as locally transmitted through fields [21,45,46]. Later, Hertz used Maxwell's results to measure electromagnetic-wave propagation over large distances [47].

But is Weber's theory correct? The answer is not straightforward—just as it is not easy to determine whether Newton's gravitational theory for planetary motion or the Schrödinger equation for quantum systems are correct. One may say that all physical theories (classical or quantum) are approximations, valid only within certain experimental scenarios. Of course, the instantaneous action at a distance predicted by Weber's formulation is incompatible with relativity (as is also the case for Newton's law and the Schrödinger equation). Is this relativistic limitation enough to classify Weber's framework as an incorrect theory? Would we reach the same conclusion for Newton's law? Or for the original Coulomb and Gauss laws? Weber's theory is valid whenever the finite propagation speed of interactions among particles does not play a significant role. For two electrons in a metal separated by a distance of, say, 10 nm, the retardation time is less than femtoseconds. Thus, in metals—where typical electron velocities are much lower than $c$—Weber's formalism provides a fully consistent framework for describing resistive, capacitive, and inductive effects [17,19]

Contrary to what might be assumed, the formulation of physical phenomena in terms of interactions among particles is not *outdated*. Through the works of Schwarzschild [22], Tetrode [23] and Fokker [24], Weber's original ideas evolved from his *instantaneous* action-at-a-distance [3] toward a *retarded* action-at-a-distance [15], allowing the description of any electromagnetic phenomena for any length and time scales. In modern times, under the assumption that both particles and fields are real elements of the theory (i.e., physical, not merely mathematical), the concept of the field introduced by Faraday and Maxwell has been quantized, giving rise to quantum field theory [25,26]. Despite its great empirical success, some mathematical inconsistencies remain in quantum field theory, which has led some physicists to revisit field-free formulations started by Weber of electromagnetic phenomena in the quantum regime [20]. For example, in 1941, Wheeler and Feynman [48] proposed quantum electrodynamics in which charged particles interact directly—through retarded times—without mediating fields. Feynman's later Nobel-winning work on the path-integral formulation of quantum electrodynamics was in part inspired by this earlier research, and still considers the concept of particle paths.

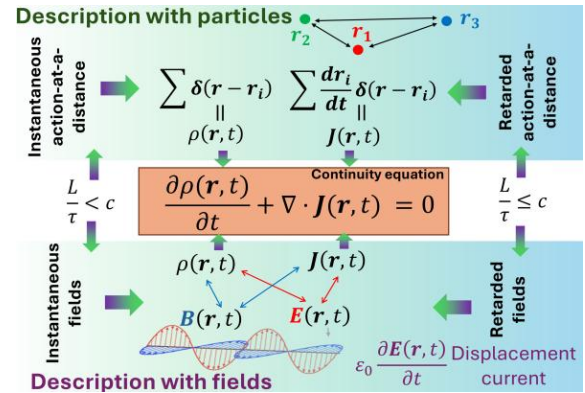

Fig. 3. *In old and modern physics, there are attempts to describe nature using either particles or fields. Both approaches have their advantages and limitations, but the need for a continuity equation is always present. In contrast, the displacement current can be understood as a byproduct of the field-based description needed to satisfy the continuity equation.*

By vindicating the ideas of action-at-distance between particles to explain electromagnetism, the reader might get the wrong impression that we are simultaneously diminishing Maxwell's explanation with fields. This is not the case. In our opinion, Maxwell's field's formulation great merit over (retarded time) particle formulation lies not in correctness, but in usefulness. It is evident that computing interparticle interactions for an Avogadro number of particles is entirely impractical in many scenarios, whereas handling such many-particle scenarios becomes relatively straightforward within Maxwell's concepts of electric and magnetic fields generated by charge and current distribution (rather than by an Avogadro number of particles). In other words, in the dichotomy between explaining electrodynamics with or without fields, the utility criterion favors the field-based approach developed by Maxwell. We are simply emphasizing that the criterion of correctness does not favor the idea of fields over particles. Fig. 3 explains this viewpoint.

In any case, in the limit of low-velocity (non-relativistic) classical electrons, Weber's theory is formally equivalent to Maxwell's equations at the conceptual level in most practical scenarios [17,19].

## IV. CONTINUITY EQUATION VERSUS DISPLACEMENT CURRENT

It is clear that Kirchhoff, when discussing electromagnetic propagation in wires, could not have developed Maxwell's displacement current because he did not employ the concepts of electric and magnetic fields. The central point emphasized throughout this paper is that he also did not need to use such concept to provide a correct explanation of signal propagation in wires. As we have seen, he was able to explain resistive, capacitive and inductive phenomena without mentioning the displacement current.

On the contrary, without the displacement current, Maxwell's set of equations is incorrect. We argue in this paper that the fundamental reason for this failure is not merely the absence of the displacement current itself, but the fact that without this term, the set of Maxwell's equations does not guarantee that charge and current satisfy a conservation law for small spatial and time intervals. When the displacement current term is added, the continuity equation—and thus charge conservation—is automatically recovered within Maxwell's equations.

The new perspective advocated in this paper is that the continuity equation is a more fundamental concept than the displacement current. The displacement current is just a byproduct of a formulation of electrodynamics with fields including the continuity equation. In this regard, Maxwell's real achievement was not the introduction of the displacement current, but the correction of a flaw in his original field-based formulation which, without this term, did not ensure the validity of the continuity equation for charge and current.

We argue in this paper that the continuity equation is a mandatory requirement when discussing electromagnetic phenomena. Let us examine whether Gauss' law [49] can be considered a correct law in the absence of the continuity equation. Gauss's law, written in terms of the electric field, is

$$\nabla \cdot \boldsymbol{E}(\boldsymbol{r}, t) = \frac{\rho(\boldsymbol{r}, t)}{\varepsilon_0}, \quad (17)$$

where $\varepsilon_0$ is the vacuum permittivity $\boldsymbol{r}$ is any point in three-dimensional space. This law alone, without the continuity equation, can lead to incorrect results. For example, consider a charge density that disappears instantaneously at one point and reappears far away; the electric fields derived from Gauss' law for such a charge distribution would be unphysical. However, within the full structure of Maxwell's equations, Gauss' law is consistent, because together with the so-called Ampere-Maxwell law ensures that matter will satisfy a continuity equation.

In a similar discussion, Assis has shown that the original Ampère's law is indeed correct if one enforces the continuity equation for charge and current. In other words, Ampère's law, together with the continuity equation, naturally leads to the discovery of the displacement current (see pp. 103–107 in Ref. [17]). Here, we summarize the mathematical demonstration. In modern notation, Ampere's law expresses the force $d\boldsymbol{F}$ experienced by an element of circuit $d\boldsymbol{l_1}$ at position $\boldsymbol{r}$, carrying a current $I_1$, due to other currents circulating nearby

$$d\boldsymbol{F} = I_1 d\boldsymbol{l_1} \times \boldsymbol{B}(\boldsymbol{r}, t), \quad (18)$$

where the influence of these other currents, represented by the current density $\boldsymbol{J}(\boldsymbol{r}', t)$ at position $\boldsymbol{r}'$ is encompassed in a magnetic field at $\boldsymbol{r}$, defined as

$$\boldsymbol{B}(\boldsymbol{r}, t) = \frac{\mu_0}{4\pi} \int \boldsymbol{J}(\boldsymbol{r}', t) \times \frac{\boldsymbol{r}' - \boldsymbol{r}}{|\boldsymbol{r}' - \boldsymbol{r}|^3} d^3 r' \quad (19)$$

with $\mu_0$ the permeability of free space. If retardation effects are significant (e.g. at high velocities), the correct expression must instead be derived from the Liénard–Wiechert potentials [15,46]. For non-relativistic scenarios, a straightforward computation of the curl of the magnetic field in Eq (19) (see pages 178 and 179 in Ref. [46]) yields

$$\nabla \times \boldsymbol{B}(\boldsymbol{r},t) = \mu_0 \boldsymbol{J}(\boldsymbol{r},t) + \frac{\mu_0}{4\pi} \nabla \int \frac{\nabla' \cdot \boldsymbol{J}(\boldsymbol{r}',t)}{|\boldsymbol{r}'-\boldsymbol{r}|} d^3 r'. \quad (20)$$

Under the steady-state assumption, $\nabla' \cdot \boldsymbol{J}(\boldsymbol{r}',t) = 0$, the familiar magnetostatic relation is recovered

$$\nabla \times \boldsymbol{B}(\boldsymbol{r},t) = \mu_0 \boldsymbol{J}(\boldsymbol{r},t). \quad (21)$$

It is important to emphasize that this last equation was never written by Ampère himself, since he did not work with fields but rather with forces between currents. In other words, Eq. (21) is a version formulated by Maxwell when rewriting Ampère's law in terms of fields. The original formulation of Ampère is simply the compact expression of the force, without reference to fields, obtained by inserting the magnetic field defined in Eq. (19) into Eq. (18).

The important point is that, without the steady-state assumption, the current density in Eq. (20) is not arbitrary; it must satisfy the three-dimensional form of the continuity equation (15), here written as

$$\frac{\partial \rho(\boldsymbol{r},t)}{\partial t} + \nabla \cdot \boldsymbol{J}(\boldsymbol{r},t) = 0. \quad (22)$$

Finally, using Gauss' law in Eq. (17) to relate the charge density to the electrical field in Eq. (22), one can rewrite Eq. (20) as

$$\nabla \times \boldsymbol{B}(\boldsymbol{r},t) = \mu_0 \boldsymbol{J}(\boldsymbol{r},t) + \mu_0 \varepsilon_0 \frac{\partial \boldsymbol{E}(\boldsymbol{r},t)}{\partial t}. \quad (23)$$

This is commonly referred to in the literature as the Ampère–Maxwell law. However, it is important to emphasize that there was nothing incorrect in Ampère's original formulation in Eqs. (18) and (19). Noticed that we do not need to "invent" the displacement current $\varepsilon_0 \frac{\partial \boldsymbol{E}(\boldsymbol{r},t)}{\partial t}$ to arrive to (23); it arises naturally from combining Ampère's law with the continuity equation. The only point to note is that, without the continuity equation in (22), Eqs. (18) and (19) could yield unphysical results if one allows arbitrary ("crazy") current densities. This conclusion mirrors what we found previously for Gauss' law. In any case, unphysical charge or current densities are not allowed when using Eqs. (17) and (23) together, because it is straightforward to show that these equations together imply the continuity equation (22).

The argument that the displacement current in Eq. (23) is merely a byproduct of the continuity equation is reinforced by the fact that this displacement current can be obtained directly from Gauss' law together with the continuity equation, without invoking Ampère's law. In fact, Gauss [49] in 1835—or even Coulomb [37] in 1785—could have "discovered" the displacement current themselves, had they employed the concept of the electric field.

Let us see how simple such a discovery could have been. By combining Gauss' law in Eq. (17) with the continuity equation in Eq. (22), one obtains the expression for the total current—the sum of particle (i.e., conduction) and displacement currents—whose divergence vanishes

$$\nabla \cdot \left( \boldsymbol{J}(\boldsymbol{r},t) + \varepsilon_0 \frac{\partial \boldsymbol{E}(\boldsymbol{r},t)}{\partial t} \right) = 0. \quad (24)$$

And that's all.

The elimination of particles in Maxwell's formulation of electromagnetism was not a necessary feature of nature, but rather a choice made by Maxwell to simplify the assumptions required to explain electrodynamics, reducing the reliance on the molecular vortices of his original model [29–32]. However, once particles are removed from the discussion, the continuity equation relating charge and current—which is automatically satisfied when particles are present—is no longer guaranteed and must be explicitly reintroduced. Electromagnetism can be formulated without fields (that is, without invoking the displacement current), but it cannot be formulated without the continuity equation for matter [50].

At this point, it should be noticed that, although the displacement current is not a fundamental concept, it remains extremely useful in many practical applications. It allows one to define a total current—the sum of conduction and displacement currents—whose divergence is zero everywhere, at all times and under all conditions. See Eq. (24). This implies that the total current entering any region of a circuit equals the total current leaving it (a property not true for conduction current alone) [51,52]. This feature has enormous practical value: in electronic devices, for example, the current predicted in the active region must equal the current measured by an ammeter located elsewhere on any time scale where the equations of electrodynamics describe experiments, certainly including the time scale of gamma rays, much faster than electron transitions in chemical reactions. This equality holds precisely because the total current includes both conduction and displacement contributions.

Another strong argument strongly supporting the perspective defended in this paper is that, in quantum electrodynamics, displacement current does not require explicit treatment. On the contrary, the continuity equation must be dealt with explicitly in quantum electrodynamics because it is a mandatory requirement that ensures gauge independent equations of motion of classical or quantum systems. See the discussion in the Appendix in Ref [27]. Moreover, the continuity equation remains valid in relativistic contexts, as it is consistent with Lorentz transformations [28].

In summary, from a fundamental standpoint, the explicit display and use of the displacement current is not strictly necessary if one avoids the field formalism altogether. What cannot be avoided—regardless of the formulation—is the continuity equation linking charge and current. This requirement is universal in any theory that includes charged particles. The main argument regarding the distinct roles played by the continuity equation and displacement current in physical theories is illustrated in Fig. 3.

## VI. SUMMARY

In this paper, we have shown that Kirchhoff deserves recognition as the first to formulate the telegrapher's equations for signal propagation in wires [1,4]. Nothing in Weber's formulation diminishes Kirchhoff's contribution to his pioneering studies of charge transport in wires [3,17,19]. In other words, Weber's work provides a valid physical theory for describing electromagnetic phenomena such as resistivity, capacitance, and induction for classical electrons moving at velocities much smaller than the speed of light, as occurs in metals. Kirchhoff's achievement is eloquently summarized in Ref. [53]: "*With these papers* [our Refs. 1,2], *he carried through the task he had envisioned in 1849: to derive the laws of current in closed circuits from Weber's fundamental law of electric action.*"

The apparent paradox of how electromagnetic propagation can be deduced without the displacement current is resolved in this paper by showing that the displacement current is a less fundamental concept than the continuity equation: electrodynamics can be formulated without invoking the displacement current (i.e., without fields), but electrical and magnetic phenomena cannot be correctly described without the continuity equation. We explicitly note that the *instantaneous* action-at-a-distance developed by Weber in the 19th century, was later extended by Schwarzschild [22], Tetrode [23], Fokker [24], Wheeler, Feynman [15], and others, in the 20th century, into the so-called *retarded* action-at-a-distance, which is valid for describing electromagnetism across all time and length scales.

As a byproduct of the present perspective, one can clarify the specific nature of Maxwell's achievement. Inspired by Faraday and his lines of force, Maxwell chose to develop electrodynamics in terms of electric and magnetic fields, thereby avoiding the need to discuss particle dynamics. When replacing the forces between particles with the more abstract concepts of fields, the inherent continuity equation provided by particle dynamics was invisible in the process. When Maxwell introduced the displacement current in his field-based formalism, what he effectively did—without explicitly realizing it [29-32]—was to impose a restriction on the kinds of currents and charges that nature considers capable of generating electromagnetic phenomena, that is, to restore the continuity equation that particle formalisms inherently satisfy.

From this new perspective, when it is said that Ampère's law is "incorrect" because it lacks the displacement current term, what should actually be stated is that Ampère's law, when written in field form, does not automatically satisfy the continuity equation, and thus there is no guarantee that it yields correct results. However, when Ampère's law is combined with the continuity equation, it becomes a correct law. Thus, Ampère's law itself is not wrong; rather, it is its field-only formulation, without the continuity equation, that is incomplete. We have seen that the same happens to Gauss' law with and without the continuity equation.

We also emphasize that, in our view, Maxwell's great merit in formulating electromagnetic phenomena in terms of fields—rather than through (instantaneous or retarded) action-at-a-distance—lies not in correctness but in usefulness. Maxwell's field-based formalism, which relates fields to charges and currents, is far more practical and broadly applicable than the particle-based force formalism. This explains why Maxwell's approach has become far more popular. There is significant merit in formulating physical theories that are not only correct but also useful.

A profound lesson emerges from the historical discussion of the continuity equation in the development of electromagnetism. Kirchhoff was inspired by Weber's formalism, but he did not model the full dynamics of particles. Instead, he was the first to impose a continuity equation for charge and current. Kirchhoff's formulation of the continuity equation might appear simple—it merely states that particles do not vanish at one point and reappear instantaneously elsewhere. However, the true significance of the continuity equation is revealed by modern formulations of electrodynamics. All of them rely on it, even those dealing with quantum matter and quantum light. In our opinion, this is a second, often overlooked, major contribution of Kirchhoff to the development of electromagnetic theory: charge and current cannot be chosen arbitrarily; they must satisfy the continuity equation given in Eq. (22). In other words, in any physical formalism developed by humans, the continuity equation must be enforced as a reminder of the fundamentally atomic nature of matter [54,55].

Why does this 19th-century 'dinosaur'—the continuity equation introduced by Kirchhoff—persist in modern quantum theories? What does this persistence indicate? One possible explanation, though not the only one, is that even in the quantum regime the notion of particles with well-defined positions—which naturally satisfy the continuity equation—remains a useful framework for discussing quantum electromagnetic phenomena [56–64].

accelerated charges (for example, electrons in the Sun), and their detection likewise involves interactions with charges (for example, electrons on Earth). Thus, electromagnetic radiation in free space is fundamentally linked to the dynamics of charged particles outside the vacuum. Of course, one may describe the process as a particle in the Sun generating an electromagnetic field that propagates through vacuum and reaches another particle on Earth. However, in the action-at-a-distance picture, the same process can be viewed equivalently as an emitter particle in the Sun interacting, after a (retarded) time, with detector particles on Earth. In both descriptions, the essential effect is the same: the motion of one particle causes the motion of another (e.g., the pointer of the measuring apparatus), regardless of the distance between them [15,22,23,24].